\documentclass[12pt]{iopart}

\usepackage{iopams} 

\usepackage{graphicx}
\usepackage{bm}

\newcommand{\rmc}{\mathrm{c}}
\newcommand{\rmH}{\mathrm{H}}

\providecommand{\openone}{\leavevmode\hbox{\small1\kern-3.8pt\normalsize1}}
\newcommand{\ie}{\textit{i.e.} }

\newcommand{\unam}{Universidad Nacional Aut\'onoma de M\'exico, M\'exico}
\newcommand{\icf}{Instituto de Ciencias F\'{\i}sicas, \unam}
\newcommand{\cic}{Centro Internacional de Ciencias, Cuernavaca, M\'exico}
\newcommand{\uPotsdam}{Institut f\"ur Physik und Astronomie, University of Potsdam, 14476 Potsdam, Germany}
\newcommand{\uDuisburg}{Fachbereich Physik, Universit\"at Duisburg--Essen, 
	D-47057 Duisburg, Germany}

\newcommand{\uGuadalajaranew}{Departamento de F\'\i sica, Universidad de Guadalajara,
  Blvd. Marcelino Garc\'\i a Barragan y Calzada Ol\'\i mpica, Guadalajara
  C.P. 44840, Jal\'\i sco, M\' exico.}

\newcommand{\mcH}{\mathcal{H}}

\def\>{\rangle}
\def\<{\langle}

\newcommand{\iint}{\int \hspace{-0.25cm} \int}

\newcommand{\cHc}{\mcH_{\rmc}}
\newcommand{\cHe}{\mcH_\rme}
\newcommand{\Hc}{H_\rmc}
\newcommand{\He}{H_\rme}

\newcommand{\tre}{\tr_\rme}
\newcommand{\trc}{\tr_\rmc}
\newcommand{\One}{\openone}

\newcommand{\la}{\langle}
\newcommand{\ra}{\rangle}
\newcommand{\lla}{\left\langle}
\newcommand{\rra}{\right\rangle}

\graphicspath{{eps/}}
\begin{document}
\title{A random matrix theory of decoherence}

\author{T Gorin$^{1}$, C Pineda$^{2,3,4}$, H Kohler$^{5}$, 
 and T H Seligman$^{3,4}$}
\address{$^1$ \uGuadalajaranew}
\address{$^2$ \uPotsdam}
\address{$^3$ \icf}
\address{$^4$ \cic}
\address{$^5$ \uDuisburg}
\ead{thomas.gorin@red.cucei.udg.mx, carlospgmat03@gmail.com}

\date{\today}
\begin{abstract}
Random matrix theory is used to represent generic loss of coherence of a {\it
fixed} central system coupled to a {\it quantum-chaotic} environment,
represented by a random matrix ensemble, via random interactions. 
We study the average density matrix arising from the ensemble induced,
in contrast to previous studies where the average values of purity, 
concurrence, and entropy were considered;
we further discuss when one or the other approach is relevant. 
The two
approaches agree in the limit of large environments.
Analytic results for the average density matrix and its purity are presented in
linear response approximation.
The two-qubit system is analysed, mainly numerically, in more detail.
\end{abstract}

\maketitle

\section{\label{I} Introduction}

Decoherence has become a subject of increasing interest, since it is not only
related to the problems of measurement and emergence of classical
behaviour~\cite{vN55a,Zeh95a,Zur91,berryMoon} but has become one of the main
stumbling blocks for quantum information solutions.
Under most usual circumstances decay of coherence is exponential and 
this has been experimentally confirmed by
experiment~\cite{LombardiExperimentDecoherence,Bru96}.
Exponential decay of coherence typically holds
if the Heisenberg time of the environment is to long to provide a relevant time
scale, i.e. if decoherence is complete long before the Heisenberg time.  As the
global environment is always present and has, for all practical purposes,
infinite Heisenberg time, our discussion seems spurious but this is not the
case. Quantum information tasks require sufficiently low temperatures, high
vacuum, good screening and careful selection of the states used etc, such that
the global environment can be neglected to first approximation. We are left
with a near environment consisting of only a few degrees of freedom which are
involved in the control of the operations of the quantum gates. This near
environment may have very specific properties, but is often governed by
dynamics that might go for ``quantum chaotic''. In this paper we provide a
generic description of decoherence which also covers such a situation.

This requires modelling of the environment, while avoiding a
complete microscopic description. 
 More specifically, we 
describe a generic environment with "quantum chaotic" dynamics by random 
matrix theory (RMT). The first step in this direction is the pioneering paper 
of Lutz and Weidenmueller~\cite{LutWei99,Lut01} in which they propose a RMT
description that reproduces the results of a Caldeira-Leggett model.
Thus they show that RMT also works in the standard situation. We
shall see that this is closely related to choosing a very dense
spectrum in the environment, which sends the Heisenberg time to
infinity, or, more generally, where decoherence is strong, long
before the Heisenberg time of the environment is reached. While this
is very reasonable for the far environment, the near environment can
have level densities comparable to those of a central system, and
this is where RMT models can produce new results. 
 A model to study such effects was first 
proposed in~\cite{GS02b,GS03} and further developed 
in~\cite{GPSZ06,pinedaRMTshort}. These models are closely
related to one developed for fidelity decay~\cite{GPS04}, which was 
successfully tested in experiments~\cite{SGSS05,SSGS05,GSW06}. For 
decoherence, the models were compared to numerics with a spin chain
environment~\cite{pineda:012305}.

In these models purity was calculated analytically in linear
response approximation as averaged over the random matrices. For the
purpose of quantum information, this was quite sufficient, as only
high purities are needed. Yet, the beauty is, that these 
RMT models are sometimes exactly solvable by super-symmetric techniques.
An attractive example are the solutions for the decay of the
fidelity amplitude~\cite{StoSch04b,StoSch05,GKPSSZ06,StoKoh06}. Yet,
these techniques nowadays are largely limited to the calculation of
two-point functions. Purity is obviously of higher order, and thus
we should look for an amplitude. Coherences, i.e. off-diagonal
elements of the density matrix are good candidates, and their decay
is often used to describe decoherence \cite{Bru96}. This
suggests to consider the average density matrix, something that
has not been done so far.

Actually, in the entire formulation of the RMT models 
a basic question has not been addressed. If we consider the evolution of some
initial state by an ensemble of Hamiltonians we also get an ensemble of density
matrices for the central system. 
Normally, we would consider the average density matrix as the basic quantity
from which observables such as purity or concurrence should be calculated.
However, since these quantities are non-linear in the density matrix, their 
average (over the ensemble of density matrices) is different from their value
computed from the average density matrix. At the end, it is the particular 
application which determines which kind of averaging must be chosen. 
We shall discuss this in \sref{V}, but we will find in the previous sections, 
that for large environments the quantities we analyse, namely purity, entropy 
and concurrence converge to the same result for both averages.

We first formulate the random matrix model we use and
derive a few basic relations, which will be needed in subsequent sections.
Some obvious conditions the model must fulfil are presented before proceeding to
calculate the average density matrix in linear response
approximation. For the particular case of one qubit as a central system we
 outline how to obtain the exact answer with
super-symmetric theory. 
Similar results for the average density matrix are presented in
what we earlier called spectator configuration \cite{pinedalong}. This configuration has
proven very useful to describe more complex central systems
with high purity \cite{GPS-letter}. We next concentrate on two-qubit central
systems in the spectator configuration, where we display some properties of the
density matrix ensemble including statistics of the eigenvalues and some further
results on a curious relation to Werner states shown in previous papers
\cite{pinedaRMTshort, pineda:012305}. 
In \sref{V} we discuss in more details the meaning of an ensemble
of density matrices.  The discussion will
include both situations where one should use the average density matrix
to calculate non-linear functionals and others where this is not
appropriate. This will lead us to conclusions and an outlook on the relevance
of more realistic RMT environments.

\section{\label{S} Random matrix models}

We shall start with a central system that does not display any particular
structure, and thus the entire system will be bi-partite consisting simply of
this central system and an environment.  Decoherence will be viewed as the
entanglement with the environment after unitary evolution of the entire system.
As we wish to describe a near environment with quantum chaotic features, we
neglect the far environment entirely, as indeed we shall throughout this paper,
assuming that decoherence resulting from the far environment occurs on a much
longer time scale.

The Hilbert space is divided into two parts, that of the central system
$\cHc$ and that of the environment $\cHe$. The dimension of those
Hilbert spaces is denoted by $m=\dim(\cHc)$ and $N= \dim(\cHe)$,
respectively.  Eventually, the case $N\to\infty$ will be
particularly important. For a factorized initial state of the form
\begin{equation}
\varrho_0= \varrho_{\rm c} \otimes \varrho_{\rm e}\;
\label{S:initstate}\end{equation}
we solve the Heisenberg equation
\begin{equation}
\rmi\, \partial_t\; \varrho(t)= H_\lambda\, \varrho(t) - \varrho(t)\; H_\lambda \; ,
\end{equation}
where the Hamiltonian is of the form
\begin{equation}\label{eq:thehamiltonian}
H_\lambda= \Hc + \He + \lambda\; V \; ,
\end{equation}
such that $H_0= \Hc + \He$. Note that typically, we shall not require the
initial state to be pure. It may well be a product of two mixed states
$\varrho_{\rm c}$ and $\varrho_{\rm e}$. In the following step, we express the
forward evolution of the whole system in terms of echo dynamics.  Echo dynamics
means, that we consider a forward time evolution which includes the coupling
between central system and environment and a backward evolution, which is
driven by the internal dynamics of environment and central system alone. The 
latter will not modify the entanglement between the two.  This is
important, because then we may apply the linear response approximation to the
echo-dynamics which results in a much larger range of validity. A systematic
application of the interaction picture would do the same simplification.
Hence, we write:
\begin{equation}
\varrho(t)= \rme^{-\rmi H_\lambda t}\; \varrho_0\;
\rme^{\rmi H_\lambda t}\; ,
\end{equation}
where we have set $\hbar = 1$. 
In order to obtain the echo dynamics, we evolve the system backwards in time
with the Hamiltonian $H_0$:
\begin{equation}
\varrho_M(t)= M_\lambda(t)\; \varrho_0\; M_\lambda(t)^\dagger\qquad
M_\lambda(t)= \rme^{\rmi H_0 t}\; \rme^{-\rmi H_\lambda t} \; .
\end{equation}
This may be interpreted as a forward evolution with the echo-operator
$M_\lambda(t)$~\cite{GPSZ06}. 
The evolution of the central system is then given by
\begin{equation}
\varrho_{\rm c}(t)= \tre\big [\, \varrho(t)\, \big ]
= \tre\big [\, (u_{\rm c} \otimes u_{\rm e})\, \varrho_M(t)\,
(u_{\rm c}^\dagger \otimes u_{\rm e}^\dagger)\, \big ]\; ,
\end{equation}
where $u_{\rm c}= \exp(-\rmi \Hc t)$ and $u_{\rm e}= \exp(-\rmi \He t)$.
This expression for $\varrho_{\rm c}(t)$ may be further simplified:
\begin{equation}
\varrho_{\rm c}(t)= u_{\rm c}\; \tre\big [\, (\One_{\rm c} \otimes
u_{\rm e})\,
\varrho_M(t)\, (\One_{\rm c} \otimes u_{\rm e}^\dagger)\, \big ]\;
u_{\rm c}^\dagger
= u_{\rm c}\; \tilde\varrho_{\rm c}(t)\; u_{\rm c}^\dagger\; ,
%= u_{\rm c}\; \tre\big [\, \varrho_M(t)\, \big ]\; u_{\rm c}^\dagger\; ,
\label{S:cecho}\end{equation}
where $\tilde\varrho_{\rm c}(t)= \tre[\varrho_M(t)]$ is the density matrix of
the central system in the interaction picture. Note that
for operators of the form $\One_{\rm c} \otimes u_{\rm e}$ the cyclic
permutations are allowed even though we are dealing with a partial trace. 
%This is an important and non-trivial result. 
\Eref{S:cecho} holds only as long as $H_0$ describes uncoupled dynamics.

\paragraph{Linear response approximation} 
In the linear response approximation, the echo-operator
$M_\lambda(t)$ reads 
\begin{equation}
M_\lambda(t) \approx \One - \rmi\lambda\; I(t) - \lambda^2 J(t)\; ,
\label{S:avM}\end{equation}
where
\begin{equation}
I(t)= \int_0^t\rmd\tau\; \tilde V(\tau)\; ,\qquad
J(t)= \int_0^t\rmd\tau\int_0^\tau\rmd\tau'\; \tilde V(\tau)\; \tilde V(\tau')
\; ,
\end{equation}
and $\tilde V(t)= \exp(\rmi H_0 t)\; V\; \exp(-\rmi H_0 t)$. 
%Equation~(\ref{S:avM}) allows to obtain the density matrix of the central
%system averaged over the coupling matrix $V$ or the environment Hamiltonian
%$\He$ or both (the averaging will be denoted by angular brackets). 
Starting from equation~(\ref{S:avM}), we obtain the density matrix of the
central system averaged over the coupling matrix $V$ (this average will be
denoted by angular brackets). Subsequently, we may also average over the
environmental Hamiltonian $H_\rme$. However, for the linear response result as
such, this is not necessary.
In the interaction picture, we therefore obtain:
\begin{eqnarray}
\fl \la\tilde\varrho_{\rm c}(t)\ra \approx
\tre\big\la\, \big (\, \One - \rmi\lambda\; I(t) - \lambda^2 J(t)\,
\big )\,
\varrho_0\, \big (\, \One + \rmi\lambda\; I(t) - \lambda^2
J(t)^\dagger\,
\big )\, \big\ra \nonumber\\
\approx \varrho_{\rm c} - \lambda^2\big (\, \la A_J\ra - \la A_I\ra\,
\big )
\; ,
\label{S:linres}\end{eqnarray}
where we have used the notation
\begin{equation}
A_J= \tre\big [\, J(t)\; \varrho_0 + \varrho_0\; J(t)^\dagger\, \big
]\; ,
\quad
A_I= \tre\big [\, I(t)\; \varrho_0\; I(t)\, \big ] \; .
\end{equation}
Note that $I(t)$ is self-adjoint, while $J(t)$ is not. According to 
\Eref{S:cecho}, this yields for the time evolution of the central 
system:
\begin{equation}
\la\varrho_{\rm c}(t)\ra = u_{\rm c}(t)\; \la\tilde\varrho_{\rm c}(t)\ra\;
u_{\rm c}(t)^\dagger \; .
\label{S:localI}\end{equation}
The averaging procedure itself as well as the different possible statistical
assumptions, will be discussed below.

\paragraph{Purity} 
We can now obtain the purity of the average density matrix of the central
system. In view of 
\eref{S:cecho}, we find for the purity of the averaged state
$\bar\varrho_{\rm c}(t)$:
\begin{equation}
P(t)= {\rm tr}\, \la\varrho_{\rm c}(t)\ra^2 = {\rm tr}\big [\, u_{\rm
c}\,
\la\tilde\varrho_{\rm c}(t)\ra\, u_{\rm c}^\dagger\, u_{\rm c}\,
\la\tilde\varrho_{\rm c}(t)\ra\, u_{\rm c}^\dagger\, \big ]
= {\rm tr}\, \la\tilde\varrho_{\rm c}(t)\ra^2 \; .
\end{equation}
The purity may be more explicitly written in terms of the average state of the
full system (in the interaction picture) as
\begin{equation}
P(t)= \trc\big (\, \tre\, \la\varrho_M(t)\ra\; \tre\, \la\varrho_M(t)\ra\, 
      \big )\; .
\end{equation}
Note that in this expression, we cannot exchange the order of the partial 
traces, since $\la\varrho_M(t)\ra$ is no longer a pure state in the product
Hilbert space $\mathcal{H}$ of central system and environment. This is the
reason why we cannot simply adopt the calculations in~\cite{pinedalong} to the 
present case. In linear response approximation we find with the help of 
\eref{S:linres}
\begin{eqnarray}
\fl P(t) \approx {\rm tr}\big (\, \varrho_{\rm c}^2\, \big )
- \lambda^2\, {\rm tr}\big ( \, \varrho_{\rm c}\, \la A_J\ra
+ \la A_J\ra\, \varrho_{\rm c} - \varrho_{\rm c}\, \la A_I\ra
- \la A_I\ra\, \varrho_{\rm c}\, \big ) \nonumber\\
\approx {\rm tr}\big (\, \varrho_{\rm c}^2\, \big ) 
- 2\lambda^2\; \big [\, {\rm tr}(\la A_J\ra\, \varrho_{\rm c})
- {\rm tr}(\la A_I\ra\, \varrho_{\rm c})\, \big ]\; .
\label{S:pulinrec}\end{eqnarray}
%In this expression the outer bracket denotes the expectation value
%with respect to $\varrho_{\rm c}$. Note that since 
Since $A_J$, $A_I$ and 
$\varrho_{\rm c}$ are all Hermitian operators, these expectation values are 
%necessarily 
real.

Note the difference to the approach chosen in
Refs.~\cite{GS02b,pineda:012305,pinedalong,GPS-letter}, where we computed the 
average purity rather then the purity of the average density matrix. This 
difference will be discussed as we proceed and taken up in detail in \sref{V}.

\paragraph{Statistical (random matrix) assumptions}
We divide the statistical assumptions employed into two parts. A basic part
which is invoked when we compute the density matrix of the central
system, and an optional part, which makes stronger assumptions on the
dynamics of the environment in order to obtain more explicit results.
\begin{itemize}
\item
We assume that $V$ is taken from the Gaussian unitary ensemble. This allows to
work in the eigenbasis of $H_0$, where $H_0$ is diagonal. The transformation
into this basis is necessarily a unitary transformation, such that the ensemble
for $V$ remains unchanged. In addition, due to the separability of 
$H_0=\He+\Hc$, the eigenbasis may be chosen as the product basis $|ij\ra$ of 
eigenstates of $\Hc$ and $\He$, respectively.
\item
The choice of $\Hc$ is quite arbitrary as long as it remains fixed. In 
contrast, we eventually assume $\He$ to be randomly chosen from an appropriate
ensemble, say one of the classical ensembles.  In that case, the two-point form 
factor $b_2(t)$~\cite{Meh91} is the only characteristic quantity of the 
spectrum. The two-point form factor may well be left unspecified. However, as 
specific examples we may consider Poisson spectra ($b_2(t)= 0$), as well as 
GOE- or GUE-spectra~\cite{Meh91}. 
\end{itemize}

\subsection{\label{SA} Average density matrix of the central state}

Here, we calculate the behaviour of the central system, when averaging
over the random-matrix coupling to the environment. The state of the central
system is 
%described by the density matrix $\bar\varrho_{\rm c}(t)$.
then described by the density matrix $\la\varrho_{\rm c}(t)\ra$. 
In view of Eqs.~(\ref{S:linres}) and~(\ref{S:localI}), we start by calculating 
the terms $\la A_J\ra$ and $\la A_I\ra$.
\begin{equation}
\la A_J\ra= \int_0^t\rmd\tau\int_0^\tau\rmd\tau'\; \tre\big [\,
\la\tilde V(\tau)\, \tilde V(\tau')\ra\; \varrho_0 + \varrho_0\;
\la\tilde V(\tau')\, \tilde V(\tau)\ra\, \big ]\; ,
\label{SA:AJ}\end{equation}
where
\begin{eqnarray}
\fl \la\tilde V(\tau)\, \tilde V(\tau')\ra = |ij\ra\;
\rme^{\rmi\, (E_{ij}\tau - E_{mn}\tau')} \lla V_{ij,kl}\, V_{kl,mn}\rra
\rme^{-\rmi\, E_{kl} (\tau -\tau')}\; \la mn| \nonumber\\
= |ij\ra\; \sum_{kl} \rme^{-\rmi\, (E_{kl} - E_{ij})(\tau-\tau')}\; \la ij|
= \bm{C}_{\rm c}(\tau-\tau') \otimes \bm{C}_{\rm e}(\tau-\tau')\; .
\label{SA:VVp}\end{eqnarray}
We use the implicit summation convention for indices appearing more than
once. The diagonal matrices $\bm{C}_{\rm c}$ and $\bm{C}_{\rm e}$ are defined
in analogy to Ref.~\cite{pinedalong} with the only difference that we use the
subscript ${\rm c}$ instead of $1$ to refer to the central system:
\begin{equation}
\fl \bm{C}_{\rm c}(\tau)= \sum_{ik} |i\ra\; \rme^{-\rmi\, (E_k-E_i) \tau}\;
\la i|
\qquad
\bm{C}_{\rm e}(\tau)= \sum_{jl} |j\ra\; \rme^{-\rmi\, (E_l-E_j) \tau}\;
\la j| \; .
\end{equation}
The separable result in \eref{SA:VVp} is due to the choice of the GUE as
random matrix ensemble for the coupling. In the GOE case, we would obtain an
additional term, which breaks the separability. 
\begin{equation}
\la A_J\ra= \int_0^t\rmd\tau\int_0^\tau\rmd\tau'\; \big [\,
C_{\rm e}(\tau-\tau')\; \bm{C}_{\rm c}(\tau-\tau')\; \varrho_{\rm c}
+ \rm{herm.}\;\rm{conjg.} \, \big ]\; ,
\label{SA:AJ2}\end{equation}
where $C_x(\tau)= {\rm tr}_x[\bm{C}_x(\tau)\, \varrho_{\rm x}]$ for
$x= {\rm c}, {\rm e}$, again in full analogy to Ref.~\cite{pinedalong}.
For the second term $\la A_I\ra$, we obtain
\begin{equation}
\la A_I\ra= \iint_0^t\rmd\tau\rmd\tau'\; \tre\big [\,
\la\tilde V(\tau)\; \varrho_0\; \tilde V(\tau')\ra\, \big ]\; ,
\label{SA:AI}\end{equation}
where
\begin{eqnarray}
\fl \tre\big [\, \la\tilde V(\tau)\, \varrho_0\, \tilde V(\tau')\ra\, \big ]
= |i\ra\; \rme^{\rmi\, (E_{ij}-E_{kl})\tau}\;
   \la V_{ij,kl}\, V_{mn,pj}\ra \; \varrho^{\rm c}_{km}\; 
   \varrho^{\rm e}_{ln}\; \rme^{\rmi\, (E_{mn}-E_{pj})\tau'}\; \la p|
\nonumber\\
 = |i\ra\; \rme^{\rmi\, (E_{ij}-E_{kl})\tau}\;
   \varrho^{\rm c}_{km}\; \varrho^{\rm e}_{ln}\; 
   \delta_{ip}\delta_{km}\delta_{ln}\;
   \rme^{\rmi\, (E_{mn}-E_{pj})\tau'}\; \la p| \nonumber\\
 = |i\ra\; \rme^{\rmi\, (E_{ij}-E_{kl})\tau}\;
\varrho^{\rm c}_{kk}\; \varrho^{\rm e}_{ll}\;
\rme^{\rmi\, (E_{kl}-E_{ij})\tau'}\; \la i| \nonumber\\
 = C_{\rm e}(\tau'-\tau)\; |i\ra\; \rme^{-\rmi\, (E_k-E_i)(\tau-\tau')}\; 
   \varrho^{\rm c}_{kk}\; \la i| \; .
\label{eq:laquees}\end{eqnarray}
This expression is a weighted version of $\bm{C}_{\rm c}(t)$, with weights 
given by the diagonal elements of $\varrho_{\rm c}$. It may be understood as a 
mapping of the density matrix $\varrho_{\rm c}$ onto a diagonal matrix.
It will be denoted by
\begin{equation}
\bm{\mathcal{C}}_{\rm c}(t)\; :\quad \varrho_{\rm c}\; \mapsto \;
   \bm{\mathcal{C}}_{\rm c}[\varrho_{\rm c}](t)= \sum_{ik} |i\ra\; 
   \rme^{-\rmi\, (E_k-E_i) t}\; \varrho^{\rm c}_{kk}\; \la i| \; .
\end{equation}
We will encounter a generalization of $\bm{\mathcal{C}}_{\rm c}(t)$ again in 
the following section. Note that
\begin{equation}
{\rm tr}\; \bm{\mathcal{C}}_{\rm c}[\varrho_{\rm c}](t)= C_{\rm c}(-t)\; .
\label{trbmmcC}\end{equation}
We may write:
\begin{equation}
\la A_I\ra= \iint_0^t\rmd\tau\rmd\tau'\; C_\rme(\tau'-\tau)\;
\bm{C}_{\rm c}[\varrho_{\rm c}](\tau-\tau')\; .
\label{SA:AI2}\end{equation}
Note that the double integral  $\la A_I\ra$ is Hermitian though the integrand 
is not.
We thus have obtained the relevant quantities for computing
the average density matrix in terms of correlation functions. In order to
obtain an explicit expression, we can subsequently average over the 
environmental Hamiltonian also.

\subsection{\label{SM} Master equation}

%In order to derive the master equation, we assume that the time scale for
%decoherence is much shorter than the Heisenberg time $\tau_H$ of the
%environment. We then approximate the spectral correlation function in the
%environment with a simple delta function:
%\begin{equation}
%C_{\rm e}(t) \to \delta(t/\tau_H)\; .
%\label{SM:assumeFGR}\end{equation}
%In the case of the random matrix environment 
%$C_{\rm e}(t) \to \bar C_{\rm e}(t)$, which contains, apart from the 
%delta function $\delta(t/\tau_H)$, also a smooth part, which becomes important
%at times of the order of $\tau_H$. For the precise form of $\bar C_{\rm e}(t)$
%see~\cite{pinedalong}.

%In what follows, we will use \eref{S:linres} to obtain a relation for
%$\tilde\varrho_{\rm c}(t)$ at short times. We will then differentiate this
%relation to obtain the desired master equation. In doing so, we assume that the
%state $\varrho(t)$ of the full system (central system and environment) remains
%a product state for all times. One common argument is that $\varrho_{\rm e}$ 
%does not change in the course of time because the environment is much bigger 
%than the central system~\cite{elementsQO}.

In order to arrive at the desired master equation, we will make two 
approximations. First, we assume that the time scale for
decoherence is much shorter than the Heisenberg time $\tau_H$ of the
environment. We then approximate the spectral correlation function in the
environment with a simple delta function:
\begin{equation}
C_{\rm e}(t) \to \delta(t/\tau_H)\; .
\label{SM:assumeFGR}\end{equation}
In the case of the random matrix environment
$C_{\rm e}(t) \to \bar C_{\rm e}(t)$, the latter also contains a smooth part,
which becomes important at times of the order of $\tau_H$ and beyond; for the 
precise form of $\bar C_{\rm e}(t)$ see~\cite{pinedalong}. Second, we assume 
that the state $\varrho(t)$ of the 
full system (central system and environment) remains a product state throughout 
the whole evolution. One common argument is that $\varrho_{\rm e}$ does not 
change in the course of time because the environment is much bigger than the 
central system~\cite{elementsQO}. Together, these approximations are consistent 
with a Markov approximation. We then use the linear response result to advance 
the density matrix $\la\varrho_{\rm c}(t)\ra$ for a time step, which should be
small compared to the decoherence time but large compared to the mixing time in 
the environment. We thereby obtain a closed differential equation for 
$\la\varrho_{\rm c}(t)\ra$, the desired master equation.

%In what follows we only deal with quantities related to the central system,
%and 
As before,
we restrict the discussion to the GUE coupling. Note though that a GOE
coupling can be treated along the same lines and would yield precisely the
same result. In this respect, on the level of a master equation approach, the
evolution of the central system does not distinguish between systems, where the
coupling breaks the time-reversal symmetry and others that do not. 
%This is completely analogous to
This is a general feature of the Fermi golden rule approximation.
% -- see e.g. weak localization correction. 
With \eref{SM:assumeFGR} and due to
$C_{\rm c}(0)= m\, \One_{\rm c}$, we find:
\begin{eqnarray}
\fl \la A_J\ra = \int_0^t\rmd\tau \int_0^\tau\rmd\tau'\; 
   \delta\left( \frac{\tau-\tau'}{\tau_H}\right)\;
   \big [\, \bm{C}_{\rm c}(\tau-\tau')\; \varrho_{\rm c}
 + \rm{herm.}\; \rm{conjg.}\, \big ]\nonumber\\
 = \frac{\tau_H^2}{2}\int_0^{t/\tau_H}\rmd x\; 
\big [\, \bm{C}_{\rm c}(0)\; \varrho_{\rm c} + {\rm herm.}\; {\rm conjg.}\, 
   \big ] = m\; t\; \tau_H\; \varrho_{\rm c} \label{SM:AJ}\; , \\
\fl \la A_I\ra = \int_0^t\rmd\tau\int_0^t\rmd\tau'\;
   \delta\left( \frac{\tau-\tau'}{\tau_H}\right)\; 
   \bm{\mathcal{C}}_{\rm c}[\varrho_{\rm c}](\tau-\tau') 
 = \tau_H^2\int_0^{t/\tau_H}\rmd x\; \One_{\rm c}
 = \tau_H\; t\; \One_{\rm c}\; , 
\label{SM:AI}\end{eqnarray}
where we have used the fact that 
$\bm{\mathcal{C}}_{\rm c}[\varrho_{\rm c}](0)= \One_{\rm c}$.
With these two results and \eref{S:linres}, we obtain for the derivative 
of $\la\tilde\varrho_{\rm c}(t)\ra$ at $t=0$:
\begin{equation}
\frac{\rmd}{\rmd\, t}\; \la\tilde\varrho_{\rm c}(t)\ra = -\lambda^2\; \big (\,
m\, \tau_H\, \varrho_{\rm c} - \tau_H\, \One_c\, \big )\; .
\end{equation}
Since we are working in the vicinity of $t=0$, we may replace $\varrho_{\rm c}$ 
with $\la\tilde\varrho_{\rm c}(0)\ra$ on the right hand side and thereby close 
the above relation:
\begin{equation}
\frac{\rmd}{\rmd\, t}\; \la\tilde\varrho_{\rm c}(t)\ra 
 = -m\,\tau_H\,\lambda^2\; 
   \big (\, \la\tilde\varrho_{\rm c}(t)\ra - m^{-1}\, \One_c\, \big )\; .
\label{SM:Meq1}\end{equation}
We therefore obtain as a closed evolution equation for the density matrix of 
the central system:
\begin{equation}
\frac{\rmd}{\rmd\, t}\; \la\varrho_{\rm c}(t)\ra 
 = -\rmi\; [\Hc ,\, \la\varrho_{\rm c}(t)\ra] - m\,\tau_H\,\lambda^2\; 
   \big (\, \la\varrho_{\rm c}(t)\ra - m^{-1}\, \One_c\, \big)\; ,
\label{SM:Meq2}\end{equation}
which constitutes the desired master equation~\cite{elementsQO}.

\subsection{\label{SP} Purity of the average density matrix}

We calculate the purity on the basis of \eref{S:pulinrec}. That implies
the calculation of two terms, ${\rm tr}(\la A_J\ra \varrho_{\rm c})$ and
${\rm tr}(\la A_I\ra \varrho_{\rm c})$. For the first term we obtain from 
\eref{SA:AJ2}:
\begin{eqnarray}
\fl {\rm tr}(\la A_J\ra \varrho_{\rm c}) 
 = \int_0^t\rmd\tau\int_0^\tau\rmd\tau'\;
{\rm tr}\big [\, C_{\rm e}(\tau-\tau')\; \bm{C}_{\rm c}(\tau-\tau')\;
\varrho_{\rm c}^2 + \rm{herm.}\; \rm{conjg.}\, \big ] \nonumber\\
= \iint_0^t\rmd\tau\rmd\tau'\; C_{\rm e}(\tau-\tau')\; {\rm tr}\big [\,
   \bm{C}_{\rm c}(\tau-\tau')\, \varrho_{\rm c}^2\, \big ] \; .
\end{eqnarray}
If $\varrho_{\rm c}$ is a pure state, $\varrho_{\rm c}^2 = \varrho_{\rm
c}$ and
\begin{equation}
{\rm tr}(\la A_J\ra \varrho_{\rm c}) = \iint_0^t\rmd\tau\rmd\tau'\;
C_{\rm e}(\tau-\tau')\; C_{\rm c}(\tau-\tau') \; .
\end{equation}
Similarly, we obtain:
\begin{eqnarray}
\fl {\rm tr}(\la A_I\ra \varrho_{\rm c}) = \iint_0^t\rmd\tau\rmd\tau'\;
   C_{\rm e}(\tau-\tau')\; {\rm tr}\big[\,
   \bm{C}_{\rm c}[\varrho_{\rm c}](\tau-\tau')\, \varrho_{\rm c}\, \big ]
\nonumber\\
= \iint_0^t\rmd\tau\rmd\tau'\; C_{\rm e}(\tau'-\tau)\; S_{\rm c}(\tau-\tau')
\; ,
\end{eqnarray}
where
\begin{equation}
S_{\rm c}(t) = \sum_{ik} \rme^{-\rmi (E_k-E_i) t}\; 
   \varrho^{\rm c}_{kk}\; \varrho^{\rm c}_{ii} \; .
\label{SP:defSc}\end{equation}
If $\varrho_{\rm c}$ is a pure state, the quantity $S_{\rm c}(\tau-\tau')$
agrees with that defined in Ref.~\cite{pinedalong}. Overall, we obtain
\begin{equation}
\fl
P(t)\approx 1 - 2\lambda^2\int_0^t\rmd\tau\int_0^t\rmd\tau'\; \big [\,
C_{\rm e}(\tau-\tau')\; C_{\rm c}(\tau-\tau')
- C_{\rm e}(\tau'-\tau)\; S_{\rm c}(\tau-\tau')\, \big ] \; .
\label{SP:Pres}\end{equation}
For a large environment ($N\to\infty$) and for a pure state $\varrho_{\rm c}$, 
this result agrees with the average purity calculated in 
Ref.~\cite{pinedalong}. For finite $N$, their difference is calculated in 
Appendix A. Note that neither result requires any averaging over the 
dynamics in the environment. The averaging over the coupling is sufficient.

\subsection{Towards exact expressions for the averaged Density matrix}

For small dimensions of the central system exact expressions for the ensemble 
averaged entries of the reduced density matrix can in principle be obtained by
methods similar to the ones used for the calculation of fidelity decay
\cite{GKPSSZ06,StoKoh06,correlationsRMT}.  For simplicity lets assume the
central system to be a two--level system (i.e. one qubit),
initially pure.  We write
\begin{equation}
\rho_{\rm c}(t) \ =\ \sum_{n=0}^3 \rho_{{\rm c}n}(t)\sigma_n \ , 
\end{equation}
where $\sigma_i$, $i=1,2,3$ are Pauli--matrices and $\sigma_0$ is the unit 
matrix. Then the ensemble average of $\rho_{{\rm c}n}(t) $ takes the form
 \begin{equation}
\overline{\rho_{{\rm c}n}(t)} = 
   \tr\left[ \overline{
      \rme^{-\rmi H_\lambda t}(\rho_{\rm c}\otimes \rho_{\rm e}) \;
      \rme^{\rmi H_\lambda t} (\sigma_n\otimes1_N)}\right] \; .
\end{equation}
We use an over-line to denote the ensemble average, because here, the average
is not only over the coupling but also over the environmental Hamiltonian and
even over initial states. Due to the latter, the above expression further 
simplifies to
\begin{eqnarray}
\label{fidgen}
\overline{\rho_n(t)} = \sum_{m=0}^3\rho_{{\rm c}m}(0)\; f_{nm}(t) \; , \qquad
 f_{nm}(t)= \frac{1}{2N} \tr\left[ \overline{ \rme^{-\rmi H_\lambda t}\; 
      \rme^{\rmi \sigma_m H_\lambda \sigma_n t}}\right] \; .
\end{eqnarray}
The quantities $f_{nm}(t)$  are formally  average fidelity amplitudes or 
--equivalently-- the ensemble averages of the trace of echo operators. The 
forward time evolution is given by the $2N\times2N$ matrix 
$\sigma_m H_\lambda \sigma_n$ and the backward time evolution is propagated by 
$H_\lambda$. Fidelity amplitudes have been calculated exactly for several 
different random matrix ensembles using supersymmetry. The calculation of 
$f_{nm}(t)$ is therefore amenable to an exact treatment using the techniques 
of~\cite{StoKoh06,guhr98randomfull,Efetov:1983xg}. We shall present the result 
in a future paper.

\section{\label{M} Decoherence within the spectator model}

\paragraph{Dynamical model} In the spectator model, the central system is
divided into two parts 
$\mathcal{H}_{\rm c}= \mathcal{H}_1 \otimes \mathcal{H}_2$, where only one
part is coupled to the environment (in our case this will be $\mathcal{H}_1$).
The Hilbert space of the environment is as before $\mathcal{H}_{\rm e}$. The 
dimension of the different spaces are: $m_1= \dim(\mathcal{H}_1)$, 
$m_2= \dim(\mathcal{H}_2)$, and $N= \dim(\mathcal{H}_{\rm e})$. The dimension 
of the total Hilbert space is $N\, m$, where $m= m_1\, m_2$. The Hamiltonian 
reads
\begin{equation}
H= H_1 + H_2 + H_e + V_{1,\rm e} \otimes \One_2\; .
\end{equation}
For an initial state $\varrho_{\rm c}\otimes\varrho_{\rm e}$ this means
only $\mathcal{H}_1$ and $\mathcal{H}_{\rm e}$ are coupled dynamically, whereas
$\mathcal{H}_2$ may affect the dynamics of the whole system only due to an
initial entanglement, \ie a non-separable $\varrho_{\rm c}$.

\paragraph{Linear response approximation}
The spectator model may be treated along the same lines as the non-structured
model considered in \sref{S}. Thus we may use Eqs.~(\ref{S:linres}) 
and~(\ref{S:localI}) without change, such that the principal task is to compute 
$\la A_J\ra$ and $\la A_I\ra$ on the basis of \eref{SA:AJ} 
and~(\ref{SA:AI}), respectively. Hence, it is only the coupling matrix in the 
interaction picture together with its second moment, what we need to calculate 
anew. To that end we introduce $u_1(t)= \exp(-\rmi H_1 t)$, 
$u_2(t)= \exp(-\rmi H_2 t)$ and denote with $V_{1,{\rm e}}$ the coupling matrix 
acting on the Hilbert space $\mathcal{H}_1\otimes \mathcal{H}_{\rm e}$. We 
therefore obtain:
\begin{eqnarray}
\fl \tilde V_{1,{\rm e}}(t) =
u_1(t)^\dagger\otimes u_{\rm e}(t)^\dagger\otimes u_2(t)^\dagger\;
\big [\, V_{1,\rme}\otimes\One_2\, \big ]\; u_1(t)\otimes u_{\rm e}(t)\otimes
u_2(t) \nonumber\\
= \big [\, u_1(t)^\dagger\otimes u_{\rm e}(t)^\dagger\; V_{1,\rme}\;
u_1(t)\otimes u_{\rm e}(t)\, \big ] \otimes \One_2\; .
\end{eqnarray}
This result can be interpreted as the ``old'' random coupling matrix in the
interaction picture acting in the Hilbert space 
$\mathcal{H}_1\otimes \mathcal{H}_{\rm e}$ times the identity in 
$\mathcal{H}_2$. As a consequence:
\begin{eqnarray}
\la\tilde V_{1,{\rm e}}(\tau)\, \tilde V_{1,{\rm e}}(\tau')\ra =
  \bm{C}_1(\tau-\tau') \otimes \One_2 \otimes \bm{C}_\rme(\tau-\tau') \\
\la A_J\ra = \int_0^t\rmd\tau\int_0^\tau\rmd\tau'\; \big [\, C_\rme(\tau-\tau')
   \; \bm{C}_1(\tau-\tau')\otimes\One_2\; \varrho_{\rm c} + 
   \rm{herm.}\;\rm{conjg.} \, \big ]\; .
\end{eqnarray}
In the case of $\la A_I\ra$ the calculation is a bit more involved:
\begin{eqnarray}
\fl {\rm tr}_\rme\big [\, \la\tilde V_{1,\rme}(\tau)\; \varrho_0\; 
   \tilde V_{1,\rme}(\tau')\, \big ]= |ij\ra\; 
   \rme^{\rmi (E_i+E_j+E_\alpha)\tau}\; \big\la\, V^{1,\rme}_{i\alpha,k\beta}\; 
   \delta_{jl}\; \rme^{-\rmi (E_k+E_l+E_\beta)\tau}\; \varrho^{\rm c}_{kl,mn}\;
   \varrho^\rme_{\beta\gamma}\nonumber\\
\qquad\times \rme^{\rmi (E_m+E_n+E_\gamma)\tau'}\;
   V^{1,\rme}_{m\gamma,p\alpha}\,\big\ra\; \delta_{nq}\;
   \rme^{-\rmi (E_p+E_q+E_\alpha)\tau'}\; \la pq| \nonumber\\
= |ij\ra\; \rme^{-\rmi (E_k-E_i)(\tau-\tau')}\; \varrho^{\rm c}_{kj,kn}\;
   \rme^{-\rmi (E_\beta-E_\alpha)(\tau-\tau')}\; \varrho^\rme_{\beta\beta}\;
   \la in| \nonumber\\
= C_\rme(\tau'-\tau)\; \sum_{ij,kn} |ij\ra\; 
   \rme^{-\rmi (E_k-E_i)(\tau-\tau')}\; \varrho^{\rm c}_{kj,kn}\; \la in|\; .
\label{M:AI}\end{eqnarray}
This result is quite similar to the expression in \eref{eq:laquees} for the
same quantity in the non-structured case, where we had defined the
operator-mapping $\bm{\mathcal{C}}_{\rm c}(t)$. One prominent
feature of this mapping was the fact that it transformed any matrix
into a diagonal one. In the present expression this occurs only
partially, namely in $\mathcal{H}_1$, while the density matrix
remains in a sense untouched in $\mathcal{H}_2$. Thus we generalize
$\bm{\mathcal{C}}_{\rm c}(t)$, in what follows $\bm{\mathcal{C}}_1(t)$, to 
cases where its argument acts on a product space 
$\mathcal{H}_{\rm c}= \mathcal{H}_1\otimes \mathcal{H}_2$, in which case the 
operator mapping yields exactly what we obtained in \Eref{M:AI}.
\begin{equation}
\bm{\mathcal{C}}_1(t)\; :\quad \varrho_{\rm c} \;\mapsto\; 
   \bm{\mathcal{C}}_1[\varrho_{\rm c}](t)= \sum_{ij,kn} |ij\ra\;
   \rme^{-\rmi (E_k-E_i)(\tau-\tau')}\; \varrho^{\rm c}_{kj,kn}\; \la in|\; .
\end{equation}
This allows us to write:
\begin{equation}
\la A_I\ra = \iint_0^t\rmd\tau\, \rmd\tau'\; C_\rme(\tau'-\tau)\;
\bm{\mathcal{C}}_1[\varrho_{\rm c}](\tau-\tau') \; .
\end{equation}

\subsection{\label{MM} Master equation}

Following the same lines as the derivation of the master equation in
the non-structured case, we find in the Fermi golden rule limit:
\begin{equation}
\la A_J\ra = m_1\; \tau_H\; t\; \varrho_{\rm c} \; , \qquad
\la A_I\ra = 
	\tau_H\; t\; \One_1 \otimes {\rm tr}_1\, \varrho_{\rm c} \; .
\end{equation}
Therefore
\begin{equation}
\la\tilde\varrho_{\rm c}(t)\ra= \varrho_{\rm c} 
 - \lambda^2\, \tau_H\, t\; \big [\, m_1\; \varrho_{\rm c} 
 - \One_1 \otimes {\rm tr}_1\, \varrho_{\rm c}\, \big ] \; , 
\end{equation}
which may be continued in time, just as in the non-structured case, to give a
closed differential equation for $\la\tilde\varrho_{\rm c}(t)\ra$:
\begin{equation}
\frac{\rmd}{\rmd t}\; \la\tilde\varrho_{\rm c}(t)\ra
 = - m_1\, \tau_H\, \lambda^2\; \big [\, \la\tilde\varrho_{\rm c}(t)\ra 
   - m_1^{-1}\; \One_1\otimes {\rm tr}_1\, \la\tilde\varrho_{\rm c}(t)\ra\, 
   \big ] \; , 
\label{MM:maequ}\end{equation}
where $\la\tilde\varrho_{\rm c}(0)\ra= \varrho_{\rm c}$.

\paragraph{Werner state solution}
Recently, \cite{pineda:012305,pinedalong}, it has been 
observed that the average purity and the average concurrence share a one-to-one 
correspondence during the decay of an initial Bell state. This correspondence
is the same as for the one-parameter family of Werner states. Among other 
things, \cite{pineda:012305,pinedalong} consider the decoherence in 
a central two-qubit system coupled in the spectator configuration to a random 
matrix environment, and it is under these circumstances that the curious 
relationship can be observed. The master equation derived above resolves 
this puzzle, since, as we will show below, Werner states are solutions to that 
equation. The point is that the average purity (concurrence) agrees with the 
purity (concurrence) of the average density matrix in the limit of a large 
environment ($N\to\infty$). In the case of purity, this has been shown 
analytically within the linear response approximation and numerically in 
\sref{pineda}. In the case of concurrence this has been shown only 
numerically (in the same \sref{pineda}).
 
Let us consider the two-qubit system of Sec.~\ref{pineda}. We define a Werner
state as a two-qubit mixed state which may be decomposed into a maximally 
entangled pure state and the identity matrix:
\begin{equation}
\varrho_W(t)= \alpha(t)\; \frac{\One_{\rm c}}{4} + \beta(t)\; |\psi\ra\la\psi| 
\; ,
\end{equation}
where $\psi$ is the maximally entangled initial state, and 
$\alpha(t) + \beta(t) = 1$. Maximal entanglement is equivalent to the 
condition
\begin{equation}
{\rm tr}_1\, |\psi\ra\la\psi| = \frac{\One_2}{2} \quad\Leftrightarrow\quad
 {\rm tr}_2\, |\psi\ra\la\psi| = \frac{\One_1}{2}\; . 
\end{equation}
We will show that $\varrho_W(t)$ solves the master \eref{MM:maequ} for 
appropriately chosen coefficients $\alpha(t)$ and $\beta(t)$. Note that 
\eref{MM:maequ} determines the time evolution of the two-qubits in the 
interaction picture. In order to obtain $\la\varrho_{\rm c}(t)\ra$ which 
describes the time evolution of the central system in the Schr\" odinger
picture, we have to apply \eref{S:localI}. If we have some single qubit 
dynamics, that may result in $\la\varrho_{\rm c}(t)\ra$ being no longer
a Werner state. To complete the proof, we simply make the Ansatz:
\begin{equation}
\tilde\varrho_{\rm c}(t)= \varrho_W(t)\; , \qquad 
\tilde\varrho_{\rm c}(0)= |\psi\ra\la\psi| \; .
\end{equation}
Substitution into the master equation~(\ref{MM:maequ}) yields:
\begin{equation}
\dot\alpha\; \frac{\One_{\rm c}}{4} + \dot\beta\; |\psi\ra\la\psi|= 
  -2\tau_H\, \lambda^2\; \beta
   \left( |\psi\ra\la\psi| - \frac{\One_{\rm c}}{4}\right)\; .
\end{equation}
Taking the trace on either side, we find: $\dot\alpha + \dot\beta = 0$, in 
agreement with the normalisation condition above. However, projecting either 
sides on $|\psi\ra\la\psi|$, we obtain:
\begin{equation}
\fl \frac{\dot\alpha}{4} + \dot\beta = -2\tau_H\,\lambda^2\; \frac{3\beta}{4}
\quad\Rightarrow\quad \dot\beta = -2\tau_H\, \lambda^2\; \beta
\quad\Rightarrow\quad \beta(t) = \rme^{-2\tau_H\, \lambda^2\; t} \; .
\end{equation}
As announced, this determines $\alpha(t)$ and $\beta(t)$ in such a way that
the Werner state $\varrho_W(t)$ solves the master equation.

\subsection{\label{MP} Purity}

The linear response expression~(\ref{S:pulinrec}) for purity equally
holds for the present spectator model.
Thus
\begin{equation}
P(t)= {\rm tr}\, \la\tilde\varrho_{\rm c}(t)\ra^2 \approx 
   {\rm tr} \varrho_{\rm c}^2 
 - 2\lambda^2\; {\rm tr}\big [\, {\rm tr}(\la A_J\ra\varrho_{\rm c})
   - {\rm tr}(\la A_I\ra \varrho_{\rm c})\, \big ]\; ,
\end{equation}
where
\begin{eqnarray}
{\rm tr}(\la A_J\ra \varrho_{\rm c}) = \iint_0^t\rmd\tau\rmd\tau'\;
C_\rme(\tau-\tau')\; {\rm tr}\big [\, \bm{C}_1(\tau-\tau')\otimes\One_2\;
\varrho_{\rm c}^2\, \big ] \\
{\rm tr}(\la A_I\ra \varrho_{\rm c}) = \iint_0^t\rmd\tau\rmd\tau'\;
   C_\rme(\tau'-\tau)\; {\rm tr}\big [\, 
   \bm{\mathcal{C}}_1[\varrho_{\rm c}](\tau-\tau')\, \varrho_{\rm c}\, \big ]
\end{eqnarray}
If $\varrho_{\rm c}$ represents a pure state, the purity $P(t)$ in
linear response approximation depends on the initial state only via
${\rm tr}_2\, \varrho_{\rm c}$. This can be seen from:
\begin{equation}
{\rm tr}(\la A_J\ra \varrho_{\rm c}) = \iint_0^t\rmd\tau\rmd\tau'\;
C_\rme(\tau-\tau')\; {\rm tr}\big [\, \bm{C}_1(\tau-\tau')\; {\rm tr}_2\,
   \varrho_{\rm c}^2\, \big ]\; ,
\end{equation}
which is true for any density matrix $\varrho_{\rm c}$, and
\begin{equation}
{\rm tr}(\la A_I\ra \varrho_{\rm c}) = \iint_0^t\rmd\tau\rmd\tau'\;
   C_\rme(\tau'-\tau)\; {\rm tr}\big [\, \bm{\mathcal{C}}_1[{\rm tr}_2\,
   \varrho_{\rm c}](\tau-\tau')\, \big ] \; ,
\end{equation}
which is only true if $\varrho_{\rm c}$ represents a pure state.

\section{\label{pineda} Two-qubit central systems}

We shall proceed to analyse properties of the average density matrix for the
particular case of a central system consisting of two qubits. Most of the
analysis will be numerical, and thus we need a small central system for
numerical expedience. A single qubit is of interest, but choosing this case
would eliminate the option of considering entanglement within the central
system. Anyway, the one qubit case will emerge as a special case, if the
spectator is not entangled.

For the two qubit case, we consider not only the purity of the density matrix
$\rho$ but also its von Neumann entropy $S(\rho) = - \tr \rho \log \rho$. The
latter quantity is preferred by many authors as a measure of decoherence. 
We shall quantify internal entanglement with concurrence, and
study its properties in the spirit mentioned above. It is defined as
$C(\rho)=\max \{0, \lambda_1- \lambda_2- \lambda_3- \lambda_4 \}$ where
$\lambda_i$ are the eigenvalues of the matrix $\sqrt{\rho (\sigma_y \otimes
\sigma_y) \rho^* (\sigma_y \otimes \sigma_y)}$ in non-increasing order, $(^*)$
denotes complex conjugation in the computational basis, and $\sigma_y$ is a
Pauli matrix~\cite{wootters}. 

\begin{figure}
\begin{center} \includegraphics{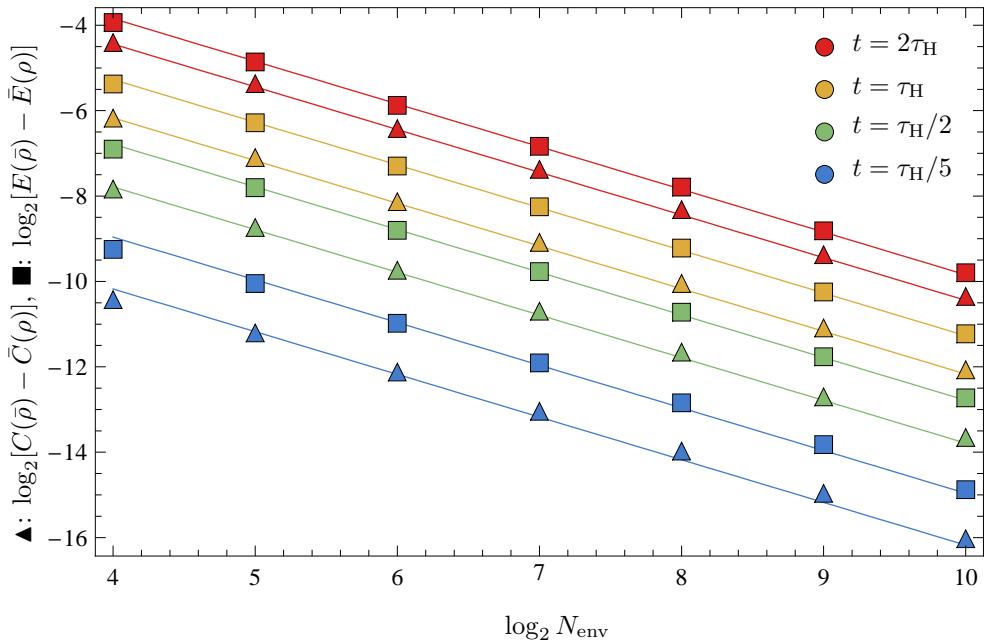} \end{center}
\caption{\label{fig:aCEvsCEa} Difference between the entropy
  (squares)/concurrence (triangles) of the average density matrix and their
  value for each density matrix, averaged over the ensemble. The colours
  indicate different times and the lines correspond to a fit with fixed slope
  of $-1$. This shows that the average density
  matrix captures the physics of the ensemble for the quantities considered,
including purity in the large $N$ limit. Here, $\lambda=0.03$, and the average values
  of purity are $\bar P(\tau_\rmH/5) \approx 0.99$, $\bar P(\tau_\rmH/2) \approx
  0.95$, $\bar P(\tau_\rmH)\approx 0.88$ and $\bar P(2 \tau_\rmH)=0.66$.}
\end{figure}

In figure \ref{fig:aCEvsCEa} we show the differences between the ensemble
averages of $C$ and $S$ as well as their value for the average density matrix
$\bar \rho$, for different times and stepwise increasing the size of the
environment. As the reader can see this difference approaches zero with
increasing environment size as $1/N_{\rm env}$. Similar results were obtained
for purity, but are not shown in a figure.  They are consistent with the
results found in \sref{S}. The upshot is then, that we see all
these very relevant quantities do not depend on the form in which the average
is taken. Whether this consequence of a concentration of measure to some
generalized delta function may be conjectured, but the demonstration thereof
will be then interesting subject of future research.

Concurrence and purity, for two non-interacting qubits coupled moderately to an
environment, display a relation which corresponds well to that of a Werner
state, while the members of the ensemble are {\it not} Werner states.  It is
thus reasonable to ask, whether the average density matrix corresponds to a
Werner state.  A 2 qubit Werner state, can be alternatively be defined as a
density matrix with three degenerate eigenvalues, whose non-degenerate
eigenstate is a maximally entangled state (i.e. a Bell state modulo local
unitary transformations). We first examine under which circumstances we have a
triple degeneracy.

\begin{figure}
\begin{center} \includegraphics{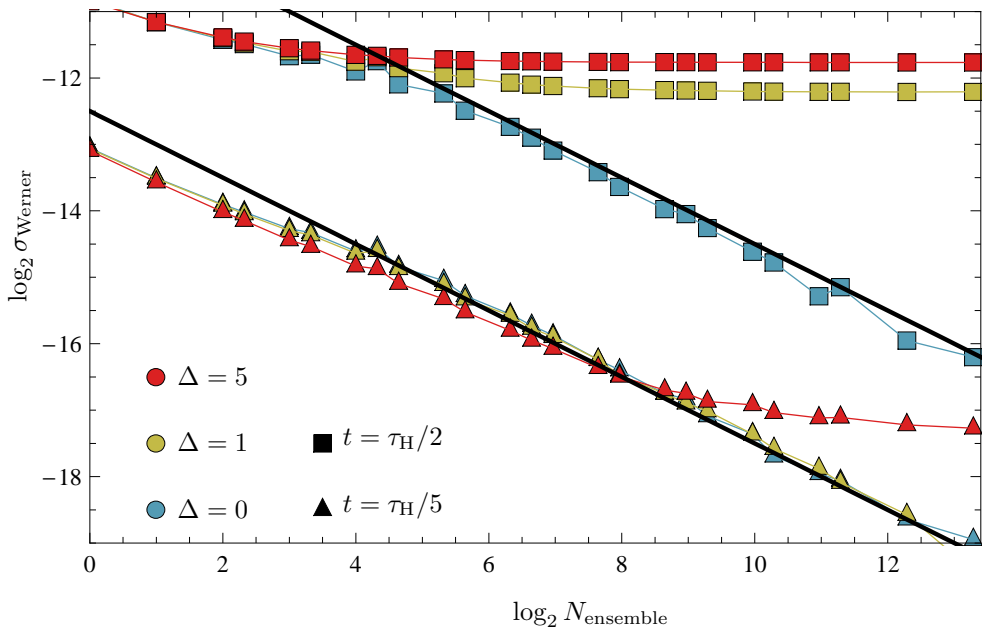} \end{center}
\caption{\label{fig:threeandtwoD} Standard deviation of the three 
closest eigenvalues of the average density matrix as a function
of the size of the ensemble (see text for details). Different
level splittings (differentiated by colours) and different times
(differentiated by symbol type) are studied. When the coupled
qubit has no internal Hamiltonian (blue symbols), the density
matrix $\rho$ has a triply degenerate eigenvalue in the large 
dimension limit. However when $\Delta>0$, this triple degeneracy is 
lifted, as can be seen from an asymptotic non-zero value. }
\end{figure}

Numerically, we construct a finite ensemble of $N_{\rm tot}$ density
matrices.  We partition it in sets of equal size $N_{\rm par}$, and
evaluate the average density matrix $\rho_{N_{\rm par}}^{(i)}$ for
each of the $N_{\rm tot}/N_{\rm par}$ members of the partition
($i=1,\cdots,N_{\rm tot}/N_{\rm par}$).  The properties of
$\rho_{N_{\rm par}}^{(i)}$ are studied. In particular we consider
the standard deviation $\sigma_{\rm Werner}$ of its three closest
eigenvalues. The average $\sigma_{\rm Werner}$ for each of the
$N_{\rm tot}/N_{\rm par}$ sets available. This quantity is plotted
in \fref{fig:threeandtwoD} for different level splitting $\Delta$,
[$H_\rmc= \Delta \sigma_z/2 $ in \eref{eq:thehamiltonian}].

Indeed, if the coupled qubit has no internal dynamics, the average
state, in the large dimension limit, is triply degenerate.  In fact $\sigma_{\rm
Werner} \to 0$ as  $1/\sqrt{N_{\rm par}}$.  If the qubit has an
internal Hamiltonian, the degeneracy is lifted: a systematic
splitting of the eigenvalues (and hence a deviation from Werner
states) happens on a time scale set by the mean level spacing of
the environment.

We also studied the concurrence of the eigenvector corresponding to the
non-degenerate eigenvalue. It remains very close to one, independently of the
concurrences of the other eigenvectors, which fluctuate  without a clear
pattern.  We find the rather surprising result, that this property of the
eigenfunction of the largest eigenvalue  persists even in cases where the
previous test shows, that we no longer deal with Werner states. Whether this
property is characteristic of the average density matrix or whether it is
common to many models will have to be analysed in future work.

%%%%%%%%%%%%%%%%

\section{\label{V} Ensembles of density matrices and average density matrix}

The very spirit of the treatment of decoherence and entanglement by RMT leads
to ensembles of density matrices: each member of the RMT ensemble (and, in our
approach, also each initial condition in the environment) induces a density
matrix $\rho_i$ via unitary dynamics followed by partial tracing.  For that
ensemble of density matrices we may consider the average density matrix $\bar
\rho= \overline{ \rho_i}$. A difference of background will lead at this point
to a natural inclination to two different approaches: Physicists with a
background in decoherence, quantum optics and quantum information will most
likely tend to use this average density matrix to calculate any quantity
desired [say $C(\bar \rho)$], while others with a background in random matrix
theory will be inclined to calculate the ensemble averages of the quantities of
interest [say $\overline {C(\rho_i)}$].  The difference between these two
approaches disappears when one studies expected values of observables as $\tr A
\bar \rho = \overline{ \tr A \rho_i}$ or, as show in this paper, for some
non-linear quantities in the limit of large dimension for the environment.
Moreover, with quantitative entanglement witnesses
\cite{entanglement-quantitative}, one can bound some of these non-linear
quantities using (linear) observables.  Yet the very concept of a relevant near
environment, important for many of the new insights obtained form an RMT
treatment, suggests to consider small environments as well.  This leads us to
discuss the appropriateness of either approach in typical applications.

One must recall, that the determination of any of these quantities is not
achievable as a single measurement. Rather we need an ensemble of ``identical''
systems. The hinge is what we mean by ``identical''. If we have no control over
the state or dynamics of the near environment, we must consider $f(\bar \rho)$
($f$ being an expectation value, or any of the non-linear quantities
discussed).  Two notes are important at this point: First, the problem
discussed here appears for any type of averaging, not only for averaging over
time evolutions. Second, for the GUE the ensemble a state average is implicit
in the ensemble average, while this is not the case for a GOE \cite{goegue},
and thus has to be performed separately.  However, if the near environment is
known and its dynamics can be controlled, one should consider
$\overline{f(\rho)}$. This can indeed be the case e.g. for an ion trap setup as
the one discussed in the introduction, which might be used to experimentally
verify our results.

The differentiation between both approaches is also relevant within quantum
information tasks.  Consider for instance Shor's algorithm. The success rate
has to be calculated from the average density matrix $\bar \rho$ resulting from
many realizations of the same experiment. For the teleportation of an unknown
state, however, it is important to know the success rate at each attempt, and
thus the quality of the process should result from an average quality for
individual density matrices corresponding to each try.

At this point, to avoid confusion, we should recall that there is the option of
obtaining a mixed density matrix without any entanglement by allowing
probabilistic variations in the unitary time evolution of the central system
itself in the absence of any significant coupling to the environment. Clearly
we then have an ensemble of pure states, whose average naturally will not be
pure.  Yet errors due to variations of the unitary evolution are usually
considered as a loss of fidelity. The purity of the mixed density matrix would
then measure an average quantity closely related to fidelity. This explains our
emphasis throughout this paper on the fact, that we need a fixed evolution for
the central system.  It also shows, that the separation of external and
internal perturbations for the study of the stability of quantum information
tasks is somewhat artificial in a practical sense, though it is very useful for
theoretical studies.

The upshot of this discussion resides in the very fact, that for many quantum
information tasks, we need repeated experiments.  The decision, which average
should be taken, is a subtle one. This emphasizes the importance of the result
indicating, that the averages coincide in many cases.

\section{\label{C} Conclusions}

We have presented a random matrix theory of decoherence and entanglement. The
need for such a theory derives from two facts. On the one hand random matrix
theory is known to provide a good generic description of properties of systems
displaying, what is often known as quantum chaos or wave chaos. Indeed this
statement is almost tautological as many authors nowadays omit any relation to
classical chaos and use the relation to statistical properties of RMT as the
definition of the latter. Fidelity decay is the other determining factor for
the stability of quantum processes, in particular for those relating to quantum
information. RMT has proven very successful in this field even in describing
experiments \cite{SGSS05,SSGS05,GSW06}, and comparison with numerics for
dynamical spin chains have also given encouraging results
\cite{pp2007,pinedalong,pinedaseligmanELAF}.  On the other hand the universal regime
of exponential decay of coherence usually considered and derived in many ways
(including RMT \cite{LutWei99,Lut01}) is basically founded in the typical
situation of very long Heisenberg times in the environment.  Intuitive
arguments based on Fermi's Golden rule make this behaviour plausible almost
independently of the model of the environment used, whenever decoherence occurs
long before this time. Yet we argued, that situations with a near environment
with fairly low level density and thus short Heisenberg times occur, and will
become standard as quantum information systems with ever better isolation from
the general environment are developed.  It is under these circumstances, that
RMT can provide the generic model to which the behaviour of specific systems
should be compared. After describing the RMT model family, proposed to a large
extent in earlier work, we proceed to analyse a point, which has set RMT models
apart from other models of decoherence.  The ensemble of evolution operators
creates for the central system an ensemble of density matrices rather than a
single density matrix. Consequently, properties such as entropy, purity and
concurrence have so far been calculated as averages over that ensemble.
However, we can also compute the average density matrix first, putting us on
equal footing with other more conventional models. The various properties
mentioned above are then determined from that single density matrix.  This has
also the great advantage to produce lower order quantities that have a fair
chance to be calculated exactly using super-symmetric techniques.  While we
have not yet achieved this goal, we have calculated the average density matrix
in linear response approximation. This as well as numerics allowed us to
compare purity, von Neumann entropy as well as concurrence of the average
density matrix to the average of these quantities over the ensemble of density
matrices.  The central finding is, that for large environments at constant
Heisenberg time (or mean level distance) the difference between the two
approaches converges to zero as the inverse of the dimension of the relevant 
Hilbert space of the environment.  This indicates, that for decoherence we
can often use the average density matrix, and thus the RMT models are really on
the same footing with usual descriptions. Yet we have to note, that, when
describing the entanglement between two smaller systems, deviations are
important and the question which average the behaviour of a single system
should be compared to depends on the particular experimental situation.  
Consistently, if we specialize to a two-qubit central system, we find that
average concurrence and the concurrence of the average density matrix will also
approach the same limit for large environments. We found the more surprising
fact, that at least for small decoherence the eigenfunction of the dominant
eigenvalue of the average density matrix remains to very good approximation a
Bell state, if the initial state was a Bell state.

We have thus shown that purity and other quantities measuring entanglement
yield the same result in the large environment limit, whether calculated from
the average density matrix or as an average over the ensemble of density
matrices.  This was done analytically for purity in the linear response regime
(\sref{S}) and numerically, for purity, von Neumann entropy and concurrence
beyond the linear response regime (\sref{pineda}).

Taking into consideration the convexity of this quantities, this might imply
that in the large dimension limit of the environment the measure for the
density matrices becomes similar to a Dirac delta in the sense
that all or a large class of convex functions could be calculated directly from
the average density matrix. However other results \cite{pinedaRMTshort} suggest
that typical states are far from the average expected state. This apparent
contradiction as well as the distribution of the density matrices as such shall
be studied in a later paper.  This work can also be readily extended
considering more realistic RMT ensembles that RMT allows more realistic models
then the classical ensembles.  In particular the two-body random ensembles may
play an important role particularly in their recent formulation for
distinguishable spins \cite{PizornRMT}.  Also maps as generic models for gates
are of possible interest, and we hope that we laid the foundation for more
research in the domain of decoherence by "small" environments, where the
Heisenberg time is shorter then or of the order of the decoherence time.

\ack
We are grateful for discussions with J.P. Paz, F. Leyvraz, T. Guhr, W.
Schleich, T. Prosen, M. Znidaric, R. Blatt, H. H{\"a}ffner, J. Eisert and D.
Sanders, and we acknowledge financial support by CONACyT project 41000F, 
PAPIIT-UNAM IN101603. CP acknowledges support from CONACyT, program Estancias
Posdoctorales y Sabaticas.  HK acknowledges support from the German research
council under grant No KO3538/1-1.

\begin{appendix}
\section{Average purity for the non-structured and the spectator model}

In this section, we review the results of~\cite{pinedalong} for the average purity 
instead of the purity of the average density matrix for the models treated in
Sec.~\ref{S} and Sec.~\ref{M}. We assume that the initial state is a product
state of the form
\begin{equation}
\varrho_0 = \varrho_{\rm c} \otimes |\psi_\rme\ra\la\psi_\rme| \; ,
\end{equation}
where the state of the central system $\varrho_{\rm c}$ may be mixed, but
the state of the environment $|\psi_\rme\ra\la\psi_\rme|$ must be pure. The
latter is a special requirement which allowed to map the spectator model onto
the non-structured model in Ref.~\cite{pinedalong}.
For the average purity it was found that
\begin{eqnarray}
\fl \la P(t)\ra= P(0) - \lambda^2\big (\, B_J - B_I\, \big )\\
\fl\qquad
B_J= 4\, {\rm Re}\, p\big [\, \la J(t)\ra\, \varrho_0 \otimes \varrho_0\, 
   \big ]\\
\fl\qquad
B_I= 2\left( p\big [\, \la I(t)\, \varrho_0\, I(t)\ra \otimes \varrho_0\, 
   \big ] - {\rm Re}\, p\big [\, \la I(t)\, \varrho_0 \otimes I(t)\, \varrho_0
   \, \big ] + p\big [\, \la I(t)\, \varrho_0 \otimes \varrho_0\, I(t)\, \big ]
 \right)\\
\fl\qquad
p[A\otimes B]= {\rm tr}_\rme\big (\, {\rm tr}_{\rm c}\, A\; 
   {\rm tr}_{\rm c}\, B\, \big )\; .
\end{eqnarray}
With Eq.~(\ref{SA:VVp}) we obtain
\begin{eqnarray}
\fl B_J= 4\, {\rm Re}\int_0^t\rmd\tau\int_0^\tau\rmd\tau'\; p\big [\, \la 
   \tilde V(\tau)\, \tilde V(\tau')\ra\; \varrho_0 \otimes \varrho_0\, \big ]
\nonumber\\
= 4\, {\rm Re}\int_0^t\rmd\tau\int_0^\tau\rmd\tau'\; p\big [\, \big (\,
   \bm{C}_{\rm c}(\tau-\tau')\otimes \bm{C}_\rme(\tau-\tau')\, \big )\; 
   \varrho_0 \otimes \varrho_0\, \big ] \nonumber\\
= 4\, {\rm Re}\int_0^t\rmd\tau\int_0^\tau\rmd\tau'\; {\rm tr}_\rme\big (\,
   C_{\rm c}(\tau-\tau')\; \bm{C}_\rme(\tau-\tau')\, \varrho_\rme\; 
   \varrho_\rme\, \big )\; .
\end{eqnarray}
Since $\varrho_\rme$ is a pure state: $\varrho_\rme^2 = \varrho_\rme$, and
because $C_{\rm x}(\tau-\tau') = C_{\rm x}(\tau'-\tau)^*$, we obtain:
\begin{equation}
\fl 
B_J= 4\, {\rm Re}\int_0^t\rmd\tau\int_0^\tau\rmd\tau'\; C_{\rm c}(\tau-\tau')\;
   C_\rme(\tau-\tau')
= 2\iint_0^t\rmd\tau\rmd\tau'\; C_{\rm c}(\tau-\tau')\; 
   C_\rme(\tau-\tau')\; .
\end{equation}
This is precisely the same quantity calculated in~\cite{pinedalong} whose final
expression is given in Eqs.~(A.8) and~(A.16). It shows that the absolute
values taken
in~\cite{pinedalong}, though correct, are not really necessary. 

To calculate the first term of $B_I$ we note that
\begin{eqnarray}
\fl \la \tilde V(\tau)\; \varrho_0\; \tilde V(\tau)\ra = |ij\ra\; 
   \rme^{-\rmi (E_{kl}-E_{ij})(\tau-\tau')}\; \la kl|\, \varrho_0\, |kl\ra\;
   \la ij| \nonumber\\
= \bm{\mathcal{C}}_{\rm c}[\varrho_{\rm c}](\tau-\tau') \otimes
      \bm{\mathcal{C}}_\rme[\varrho_\rme](\tau-\tau') \; .
\end{eqnarray}
Therefore, with Eq.~(\ref{trbmmcC}), which holds equally well in the case 
of the environment:
\begin{eqnarray}
\fl p\big [\, \la I(t)\, \varrho_0\, I(t)\ra \otimes \varrho_0\, \big ] =
   \iint_0^t\rmd\tau\rmd\tau'\; C_{\rm c}(\tau'-\tau)\;
   {\rm tr}_\rme\big [\, \bm{\mathcal{C}}_\rme[\varrho_\rme](\tau-\tau')\; 
      \varrho_\rme\, \big ] \nonumber\\
= \iint_0^t\rmd\tau\rmd\tau'\; C_{\rm c}(\tau'-\tau)\; S_\rme(\tau-\tau')\; ,
\end{eqnarray}
where $S_\rme(\tau-\tau')$ is defined in precise analogy to 
$S_{\rm c}(\tau-\tau')$ in Eq.~(\ref{SP:defSc}). Similarly, we obtain for the
following terms:
\begin{eqnarray}
p\big [\, \la I(t)\, \varrho_0\,  \otimes I(t)\, \varrho_0\ra\, \big ] =
  \iint_0^t\rmd\tau\rmd\tau'\; S_{\rm c}(\tau-\tau')\; S_\rme(\tau-\tau')\\
p\big [\, \la I(t)\, \varrho_0\,  \otimes \varrho_0\, I(t)\ra\, \big ] =
  \iint_0^t\rmd\tau\rmd\tau'\; S_{\rm c}(\tau-\tau')\; C_\rme(\tau'-\tau) \; ,
\end{eqnarray}
where we had to assume again that $\varrho_{\rm c}$ represents a pure state. 
In summary, we obtain:
\begin{eqnarray}
\fl \la P(t)\ra= 1 - 2\lambda^2\iint_0^t\rmd\tau\rmd\tau'\; \big [\, 
   C_\rme(\tau-\tau')\; C_{\rm c}(\tau-\tau') - S_\rme(\tau-\tau')\; 
   C_{\rm c}(\tau'-\tau) \nonumber\\
 +\; S_\rme(\tau-\tau')\; S_{\rm c}(\tau-\tau') 
 - C_\rme(\tau'-\tau)\; S_{\rm c}(\tau-\tau') \, \big ] \; ,
\end{eqnarray}
where we have used the fact that the double integral over 
$S_\rme(\tau-\tau')\, S_{\rm c}(\tau-\tau')$ is automatically real. The whole
result is obviously invariant with respect to an interchange of the subscripts 
$\rme$ and ${\rm c}$. This means that we could have defined $p[A\otimes B]$
equally well by first tracing $A$ and $B$ over the environment and the
resulting matrix product over the central system. Which is equivalent to the
statement that the purity of the state of the environment (after tracing over
the central system) is equal to the purity of the state of the central system
(after tracing over the environment).

If we compare the result for the average purity with Eq.~(\ref{SP:Pres}) for 
the purity of the average density matrix, we find that the difference is equal 
to
\begin{equation}
\la P(t)\ra - P(t) = 2\lambda^2\iint_0^t\rmd\tau\rmd\tau'\;
   S_\rme(\tau-\tau')\; \big [\, S_{\rm c}(\tau-\tau') - C_{\rm c}(\tau'-\tau)
      \, \big ] \; .
\end{equation}
As discussed in~\cite{pinedalong}, the function $S_\rme(\tau)$ shows a similar 
behaviour as $C_\rme(\tau)$, except for an additional factor of order $N^{-1}$.
This holds at least if the state of the environment is sufficiently 
delocalized in energy. For that reason, we expect the difference 
$\la P(t)\ra - P(t)$ to be of order $N^{-1}$.

\end{appendix}

\section*{References}
%\bibliographystyle{unsrt}
%\bibliography{paperdef,mibibliografia,amol,deco,echo,ranh,semic,stas,books,qcom}

\begin{thebibliography}{10}

\bibitem{vN55a}
von Neumann J 1955
\newblock {\em Mathematical Foundations of Quantum Mechanics}
\newblock (Princeton, University Press)

\bibitem{Zeh95a}
%H.~D. Zeh.
Zeh H D 1995
\newblock Decoherence, basic concepts and their interpretation
\newblock {\em Preprint} quant-ph/9506020

\bibitem{Zur91}
%W.H. Zurek.
Zurek W H 1991
\newblock Decoherence and the transition from quantum to classical
\newblock {\em Phys. Today} {\bf 44} 36

\bibitem{berryMoon}
%M.~V. {Berry}.
Berry M V 2001
\newblock Chaos and the semiclassical limit of quantum mechanics (is the moon
  there when somebody looks?)
\newblock in Russell R J, Clayton P, Wegter-McNelly K, and Polkinghorne J, eds.
 {\em Quantum mechanics: Scientific perspectives on Divine Action}  
 (Notre Dame, Vatican Observatory -- CTNS Publications)

\bibitem{LombardiExperimentDecoherence}
%J.~Derouard, R.~Jost, and M.~Lombardi.
Derouard J, Jost R and Lombardi M 1976
\newblock Pressure broadening of an anticrossing signal
\newblock {\em Journal de Physique Lettres} {\bf 37} 135

\bibitem{Bru96}
%M.~Brune, E.~Hagley, J.~Dreyer, X.~Ma\^{\i}tre, A.~Maali, C.~Wunderlich, J.~M.
%  Raimond, and S.~Haroche.
Brune M, Hagley E, Dreyer J, Ma\^{\i}tre X, Maali A, Wunderlich C, 
  Raimond J M and Haroche S 1996
\newblock Observing the progressive decoherence of the ``meter'' in a quantum
  measurement
\newblock {\em Phys. Rev. Lett.} {\bf 77} 4887

\bibitem{LutWei99}
%E.~Lutz and H.~A.~Weidenm\" uller.
Lutz E and Weidenm\" uller H A 1999
\newblock Universality and quantum brownian motion
\newblock {\em Physica} A {\bf 267} 354

\bibitem{Lut01}
Lutz E 2001
\newblock Random-matrix model for quantum brownian motion
\newblock {\em Physica} E {\bf 9} 369

\bibitem{GS02b}
Gorin T and Seligman T H 2002
\newblock A random matrix approach to decoherence
\newblock {\em J. Opt. B: Quant. Semiclass. Opt.} {\bf 4} S386

\bibitem{GS03}
Gorin T and Seligman T H 2003
\newblock Decoherence in chaotic and integrable systems: a random matrix
  approach
\newblock {\em Phys. Lett.} A {\bf 309} 61

\bibitem{GPSZ06}
%T.~Gorin, T.~Prosen, T.~H. Seligman, and M.~{\v Z}nidari{\v c}.
Gorin T, Prosen T, Seligman T H and {\v Z}nidari{\v c} M 2006
\newblock Dynamics of Loschmidt echoes and fidelity decay
\newblock {\em Phys. Rep.} {\bf 435} 33

\bibitem{pinedaRMTshort}
%C.~Pineda and T.~H. Seligman.
Pineda C and Seligman T H
\newblock Bell pair in a generic random matrix environment
\newblock {\em Phys. Rev.} A {\bf 75} 012106

\bibitem{GPS04}
%T.~Gorin, T.~Prosen, and T.~H. Seligman.
Gorin T, Prosen T, and Seligman T H 2004
\newblock A random matrix formulation of fidelity decay
\newblock {\em New J. Phys.} {\bf 6} 20

\bibitem{SGSS05}
%R.~Sch{\" a}fer, T.~Gorin, H.-J. St{\" o}ckmann, and T.~H. Seligman.
Sch{\" a}fer R, Gorin T, St{\" o}ckmann H-J and Seligman T H 2005
\newblock Fidelity amplitude of the scattering matrix in microwave cavities
\newblock {\em New J. Phys.} {\bf 7} 152

\bibitem{SSGS05}
%R.~Sch{\" a}fer, H.-J. St{\" o}ckmann, T.~Gorin, and T.~H. Seligman.
Sch{\" a}fer R, St{\" o}ckmann H-J, Gorin T and Seligman T H 2005
\newblock Experimental verification of fidelity decay: from perturbative to
  fermi golden rule regime
\newblock {\em Phys. Rev. Lett.} {\bf 95} 184102

\bibitem{GSW06}
%T.~Gorin, T.~H. Seligman, and R.~L. Weaver.
Gorin T, Seligman T H and Weaver R L 2006
\newblock Scattering fidelity in elastodynamics
\newblock {\em Phys. Rev.} E {\bf 73} 015202(R)

\bibitem{pineda:012305}
%C.~Pineda and T.~H. Seligman.
Pineda C and Seligman T H 2006
\newblock Evolution of pairwise entanglement in a coupled $n$-body system
\newblock {\em Phys. Rev.} A {\bf 73} 012305

\bibitem{StoSch04b}
%H.-J.~St\" ockmann and R.~Sch\" afer.
St\" ockmann H-J and Sch\" afer R 2004
\newblock Recovery of the fidelity amplitude for the Gaussian ensembles.
\newblock {\em New J. Phys.} {\bf 6} 199

\bibitem{StoSch05}
%H.-J.~St\" ockmann and R.~Sch\" afer.
St\" ockmann H-J and Sch\" afer R 2005
\newblock Fidelity recovery in chaotic systems and the Debye-Waller factor
\newblock {\em Phys. Rev. Lett.} {\bf 94} 244101

\bibitem{GKPSSZ06}
%T.~Gorin, H.~Kohler, T.~Prosen, T.~H. Seligman, H.-J.~St\" ockmann, and M.~{\v
%  Z}nidari{\v c}.
Gorin T, Kohler H, Prosen T, Seligman T H, St\" ockmann H-J and 
{\v Z}nidari{\v c} M 2006
\newblock Anomalous slow fidelity decay for symmetry breaking perturbations
\newblock {\em Phys. Rev. Lett.} {\bf 96} 244105

\bibitem{StoKoh06}
%H.-J.~St\" ockmann and H.~Kohler.
St\" ockmann H-J and Kohler H 2006
\newblock Fidelity freeze for a random matrix model with off-diagonal
  perturbation
\newblock {\em Phys. Rev.} E {\bf 73} 066212

\bibitem{pinedalong}
%C.~Pineda, T.~Gorin, and T.~H. Seligman.
Pineda C, Gorin T and Seligman T H 2007
\newblock Decoherence of two-qubit systems: a random matrix description
\newblock {\em New J. Phys.} {\bf 9} 106

\bibitem{GPS-letter}
%T.~Gorin, C.~Pineda, and T.~H. Seligman.
Gorin T, Pineda C and Seligman T H 2007
\newblock Decoherence of an $n$-qubit quantum memory
\newblock {\em Phys. Rev. Lett.} {\bf 99} 240405

\bibitem{Meh91}
%M.~L. Mehta.
Mehta M L 1991
\newblock {\em Random matrices and the statistical theory of energy levels}
\newblock (New York, Academic Press)

\bibitem{elementsQO}
%P.~Meystre and M.~Sargent III.
Meystre P and Sargent III M 1998
\newblock {\em Elements of quantum optics}
\newblock (Berlin, Springer)

\bibitem{correlationsRMT}
%H.~Kohler, I.~E. Smolyarenko, C.~Pineda, T.~Guhr, F.~Leyvraz, and T.~H.
%  Seligman.
Kohler H, Smolyarenko I E, Pineda C, Guhr T, Leyvraz F and Seligman T H 2008
\newblock Surprising relations between parametric level correlations and
  fidelity decay
\newblock {\em Phys. Rev. Lett.} {\bf 100} 190404

\bibitem{guhr98randomfull}
%T.~Guhr, A.~M\"{u}ller-Groeling, and H.~A. Weidenm\"{u}ller.
Guhr T, M\"{u}ller-Groeling A and Weidenm\"{u}ller H A 1998
\newblock Random matrix theories in quantum physics: Common concepts
\newblock {\em Phys. Rep.} {\bf 299} 189

\bibitem{Efetov:1983xg}
%K.~B Efetov.
Efetov K B 1983
\newblock {Supersymmetry and theory of disordered metals}
\newblock {\em Adv. Phys.} {\bf 32} 53

\bibitem{wootters}
%W.~K. Wootters.
Wootters W K 1998
\newblock Entanglement of formation of an arbitrary state of two qubits
\newblock {\em Phys. Rev. Lett.} {\bf 80} 2245

\bibitem{entanglement-quantitative}
%J~Eisert, F~G S L~Brand\ {a}o, and K~M~R Audenaert.
Eisert J, Brand\~{a}o F G S L and Audenaert K M R 2007
\newblock Quantitative entanglement witnesses
\newblock {\em New J. Phys.} {\bf 9} 46


\bibitem{goegue}
%L.~Kaplan, F.~Leyvraz, C.~Pineda, and T.~H. Seligman.
Kaplan L, Leyvraz F, Pineda C and Seligman T H 2007
\newblock A trivial observation on time-reversal in random matrix theory
\newblock {\em J. Phys.} A {\bf 40} F1063

\bibitem{pp2007}
%C.~Pineda and T.~Prosen.
Pineda C and Prosen T 2007
\newblock Non-universal level statistics in a chaotic quantum spin chain
\newblock {\em Phys. Rev.} E {\bf 76} 061127

\bibitem{pinedaseligmanELAF}
%C.~Pineda and T.~H. Seligman.
Pineda C and Seligman T H 2008
\newblock Random matrix models for decoherence and fidelity decay in quantum
  information systems
\newblock in {\em Latin-American School of Physics XXXVIII ELAF: Quantum
  Information and Quantum Cold Matter} (New York, Springer, AIP Conference 
  Proceedings)

\bibitem{PizornRMT}
%I.~Pizorn, T.~Prosen, S.~Mossmann, and T.~H. Seligman.
Pizorn I, Prosen T, Mossmann S and Seligman T H 2008
\newblock The two-body random spin ensemble and a new type of quantum phase
  transition
\newblock {\em New J. Phys.} {\bf 10} 023020

\end{thebibliography}

\end{document}